\documentclass[conference,letterpaper]{IEEEtran}

\usepackage[utf8]{inputenc} 
\usepackage[T1]{fontenc}
\usepackage{url}
\usepackage{ifthen}
\usepackage{cite}

\usepackage[cmex10]{amsmath} 
\usepackage{xcolor}

\usepackage{amssymb}
\newtheorem{theorem}{Theorem}
\newtheorem{remark}{Remark}

\allowdisplaybreaks
\usepackage[
  letterpaper,
  left=0.673in,
  right=0.627in,
  top=1.047in,
  bottom=0.80in
]{geometry}

\begin{document}
\title{
On Resilient and Efficient Linear Secure Aggregation in Hierarchical Federated Learning
} 


\author{
  \IEEEauthorblockN{Shudi Weng$^{*}$, Xiang Zhang$^{\dagger}$, Yizhou Zhao$^{\ddagger}$, Giuseppe Caire$^{\dagger}$, Ming Xiao$^{*}$, Mikael Skoglund$^{*}$}
  \IEEEauthorblockA{$^{*}$KTH Royal Institute of Technology, Dept. of ISE, 10044 Stockholm, Sweden, Email: \{shudiw, mingx, skoglund\}@kth.se \\
    $^{\dagger}$Technical University of Berlin, Dept. of EECS, 10623, Berlin, Germany. Email: \{xiang.zhang, caire\}@tu-berlin.de.\\
    $^{\ddagger}$Southwest University, Coll. of EIE, 400715, Chongqing, China. Email: onezhou@swu.edu.cn.
}
}

\maketitle


\begin{abstract}
In this paper, we study the fundamental limits of hierarchical secure aggregation under unreliable communication. We consider a hierarchical network where each client connects to multiple relays, and both client-to-relay and relay-to-server links are intermittent. Under this setting, we characterize the minimum communication and randomness costs required to achieve robust secure aggregation. We then propose an optimal protocol that attains these minimum costs, and establish its optimality through a matching converse proof. In addition, we introduce an improved problem formulation that bridges the gap between existing information-theoretic secure aggregation protocols and practical real-world federated learning problems.
\end{abstract}

\section{Introduction}
\subsection{Hierarchical Federated Learning (HFL)}
Federated learning (FL) has emerged as a promising solution at the network edge \cite{ren2025advances,8664630}, allowing devices to collaboratively train models while retaining datasets locally, thereby enabling efficient, privacy-preserving intelligence in large-scale edge systems.
To enhance the channel quality between clients and the server, hierarchical federated learning (HFL), which introduces intermediate relay nodes that cooperatively facilitate data exchange between edge devices and the server, can effectively mitigate the adverse effects of channel fading in edge FL systems \cite{lin2022relay,chen2023relay}. 

Nevertheless, unreliable communication remains a fundamental challenge in HFL and can significantly degrade the learning performance, including optimality and privacy preservation \cite{wang2021quantized,weng2025codingenforcedrobustsecureaggregation}. First, communication uncertainties may cause uneven client participation at the server aggregation, resulting in suboptimal training performance. Second, from a privacy perspective, unreliable links may disrupt the coordination of privacy noise (or secret keys), thereby degrading global model accuracy.

\subsection{Information-Theoretic Secure Aggregation}
Existing literature has investigated the fundamental theoretical limits of communication and secret key generation in secure aggregation (SA) with information-theoretic perfect security guarantee \cite{zhao2022information,10359136,10884617,10580953,10508398,li2025capacity,zhang2025informationtheoreticdecentralizedsecureaggregation,zhang2024optimalcommunicationkeyrate,zhang2025fundamentallimitshierarchicalsecure}.
In particular, the pioneering work of
Zhao and Sun \cite{zhao2022information} introduced an information-theoretic formulation of the problem. To ensure robust secure aggregation against random user dropouts, due to e.g., unreliable communication, they proposed a two-round protocol that mitigates communication uncertainty in key cancellation. This work established an important foundation for subsequent studies. 
Later works established perfect privacy guarantees for SA under a range of constraints, including groupwise key design \cite{10359136,10884617,10580953,10508398,li2025capacity}, user collusion \cite{li2025optimalkeyratesdecentralized}, SA in various network topologies \cite{zhang2025informationtheoreticdecentralizedsecureaggregation,zhang2024optimalcommunicationkeyrate,zhang2025fundamentallimitshierarchicalsecure,zhang2025information,li2025capacity,li2025optimal, li2026optimal}, 
and generalized security models \cite{11003190, li2025collusionresilienthierarchicalsecureaggregation,li2025optimal}, as well as combinations of these constraints. 
In particular, for HFL, \cite{zhang2024optimalcommunicationkeyrate,zhang2025fundamentallimitshierarchicalsecure} makes a substantial contribution to characterizing the role of communication redundancy in SA, while \cite{li2025collusionresilienthierarchicalsecureaggregation,xu2025hierarchical} further incorporates user collusion into HFL.
Collectively, these works substantially advance the theoretical understanding of information-theoretic SA and provide a solid theoretical foundation through a systematic expansion of its scope.

Existing works tackling unreliable communication includes e.g., pairwise keys \cite{10580953,bonawitz2016practical}, two-round communication protocol \cite{zhao2022information}, etc.
However, these approaches only guarantee partial aggregation over a subset of participating clients. As a result, the learned global model may deviate from the true global optimum, potentially causing a non-vanishing performance gap \cite{wang2021quantized}.
More importantly, existing information-theoretic SA works often overlook the non-unique mapping between real-field and finite-field summations, resulting in a mismatch between the derived fundamental limits and the practical effectiveness of the proposed schemes in real-world learning processes.


\subsection{Our Contributions}
This work aims to bridge the gaps in existing research. Our main contributions are summarized as follows.
\begin{itemize}
    \item We formulate an information-theoretic hierarchical secure aggregation (HSA) problem under unreliable communication, an important yet largely unexplored setting. To distinguish,
    \begin{itemize}
        \item The global model is always formed by \emph{all} local models despite communication intermittency, thereby avoiding the adverse effects of partial participation.
        \item We take a crucial step toward applying information-theoretic SA into practical FL by ensuring a one-to-one correspondence between finite-field and real-field summations. Our results show that existing bounds remain valid in terms of degrees of freedom (DoFs), and can be made applicable to effective learning tasks by proper modifications.
    \end{itemize}
    \item We propose an optimal, resilient, and efficient HSA scheme that achieves the minimum communication and randomness costs. 
    \item We develop a novel converse proof and show that the resulting lower bounds match the achievable rates, thereby establishing the optimal rate region.
\end{itemize}


\section{Problem Formulation} \label{sec: prob_formu}
Consider an HFL protocol, consisting of $K$ clients, $K$ relays, and a server. Each client holds a local model $\boldsymbol{\Theta}_k$ of length $L$, and a secret key $\mathbf{S}_k$. The server aims to securely compute the exact sum $\sum_{k \in [K]}\boldsymbol{\Theta}_k$ without additional information leakage about any individual local models or any partial aggregation. 
Each entry $\boldsymbol{\Theta}_k(l)\in \{0, \cdots, q-1\}$ can arbitrarily distributed. The $\{\boldsymbol{\Theta}_k\}_{k\in[K]}$ are mutually independent, and are independent of $\{\mathbf{S}_k\}_{k\in[K]}$.
\vspace{-1mm}
\begin{align}
H\left(\{\boldsymbol{\Theta}_k, \mathbf{S}_k\}_{k\in[K]}\right)=\hspace{-1mm}\sum_{k=1}^{K} \hspace{-0.5mm} H(\boldsymbol{\Theta}_k)+\hspace{-0.5mm}H\left(\{\mathbf{S}_k\}_{k\in[K]}\right). 
\end{align}
During client-to-relay communication, client $k$ sends a message $\mathbf{X}_{m, k}$ to its $d\in[K]$ associated relays in $\mathcal{R}_k$ under cyclic order,
\begin{align*}
   \mathcal{R}_k \triangleq \{\, (k-i+1)\bmod K)+1 \,\}, \quad i=1,\cdots,d.
\end{align*}
where $d$ is any integer between $1$ and $K$. Accordingly, each relay $m$ hears from clients in 
\begin{align*}
    \mathcal{U}_k \triangleq \{\, (k+i-1)\bmod K)+1 \,\}, \quad i=1,\cdots,d.
\end{align*}
The message $\mathbf{X}_{m, k}\in \mathbb{F}_p^{L_X}$ sent from client $k$ to relay $m$ is a linear function of $\boldsymbol{\Theta}_k(l), l\in [L]$ and $\mathbf{S}_{m,k}$, which is generated from $\mathbf{S}_{k}$:
\begin{align}
    \mathbf{X}_{m, k}=f_{m,k}(\{\boldsymbol{\Theta}_k(l)\}_{l\in [L]}, \mathbf{S}_{m,k}), \forall k\in [K].
    \label{eq:encode}
\end{align}
Any client-to-relay communication may fail.
Let $\mathcal{V}_1$ be the set of relays that successfully receive from all $\mathbf{X}_{m, k}$ from their associated clients in $\mathcal{U}_m$. 
In the second phase, only relays in $\mathcal{V}_1\subseteq [K]$ forward the messages $\mathbf{Y}_m\in \mathbb{F}_p^{L_Y}$ to the server, where $\mathbf{Y}_m$ is a linear function of $\{\mathbf{X}_{m, k} \}_{k\in\mathcal{U}_m}$. 
\begin{align}
    \mathbf{Y}_m = h_m( \{\mathbf{X}_{m, k} \}_{k\in\mathcal{U}_m}), \forall m\in [K].
    \label{eq: encode_Y}
\end{align}
The relay-to-server communication may also fail. Let $\mathcal{V}_2\subseteq \mathcal{V}_1$ denote the set relays from which the server successfully receives the corresponding messages $\mathbf{Y}_m$.\\ 
\textbf{Correctness.} If $\lvert \mathcal{V}_2 \rvert \geq K-s$, the server must be able to decode the finite-field sum $\oplus_{k \in [K]} \boldsymbol{\Theta}_k\in\mathbb{F}_p^L$, and a subsequent unique sum $\sum_{k\in [K]} \boldsymbol{\Theta}_k\in \{0, \cdots, K(q-1)\}^L$.
\vspace{-1mm}
\begin{align}
&H\left(\underset{k \in [K]}{\oplus} \boldsymbol{\Theta}_k \bigg\vert  \left\{\mathbf{Y}_m\right\}_{k\in\mathcal{V}_2, \lvert \mathcal{V}_2 \rvert \geq K-s} \right)=0, \label{eq:correctness1}  \\
&H\left( \sum_{k\in [K]} \boldsymbol{\Theta}_k \bigg\vert \underset{k \in [K]}{\oplus} \boldsymbol{\Theta}_k \right)=0. \label{eq:correctness2} 
\end{align}   
From \eqref{eq:correctness2}, to ensure the unique mapping from the finite-field sum to real-field sum, it is required that $p$ is a prime such that $p>K(q-1)$.
\vspace{0.2em}\\
\noindent\textbf{Security.} The security at the relay $m$ must be guaranteed regardless of the communication links:
\begin{align}
I\big( \{\boldsymbol{\Theta}_k\}_{k\in[K]} ;\{\mathbf{X}_{m,k}\}_{k\in\mathcal{U}_m} \big)=0, \forall m.
\label{eq:security_relay}   
\end{align}
The security at the server must also be guaranteed regardless of the communication links.
\begin{align}
&I\bigg( \{\boldsymbol{\Theta}_k\}_{k\in[K]} ; \left\{\mathbf{Y}_m\right\}_{k\in\mathcal{V}_2} \bigg\vert \underset{k \in [K]}{\oplus} \boldsymbol{\Theta}_k \bigg)=0, \forall  \mathcal{V}_2\subseteq\mathcal{V}_1\subseteq[K].
\label{eq:security_server}   
\end{align}
\textbf{Performance metrics.} The communication rates $R_1$ and $R_2$ characterize the amount of information conveyed by each client and each relay, respectively.
\begin{align}
    R_1\triangleq\frac{\sum_{m\in \mathcal{R}_k}H(\mathbf{X}_{m,k})}{L\log p},
    R_2\triangleq\frac{\sum_{m\in[K]} H(\mathbf{Y}_m)}{K L\log p}.
\end{align}
Suppose that $\{\mathbf{S}_k\}_{k\in[K]}$ are linearly constructed from a source randomness key $\mathbf{S}_{\sum}$. 
The individual key rate $R_S$ and the source key rate $R_{S_{\sum}}$ are defined to characterize the amount of randomness required by each local model and by all local models collectively, respectively.
\begin{align}
    R_S\triangleq\frac{H(\mathbf{S}_k)}{L\log q}, \;
    R_{S_{\sum}}\triangleq\frac{ H(\mathbf{S}_{\sum})}{ L\log q}. 
    \label{eq: rate_secret}
\end{align}
A tuple $(R_1, R_2, R_S, R_{S_{\sum}})$ is said to be achievable if there exists a SA scheme such that both correctness conditions in \eqref{eq:correctness1}, \eqref{eq:correctness2} and security conditions \eqref{eq:security_relay}, \eqref{eq:security_server} are satisfied, and the rates are smaller or equal to $R_1$, $ R_2$, $R_S$, and $R_{S_{\sum}}$, respectively.
The closure of the set of all
achievable rate tuples is called the optimal region, denoted by $\mathcal{C}^*$.

\section{Main Results}
\begin{theorem}\label{Theo: 1}
In the resilient, secure HFL protocol under unreliable communication, as described in Section~\ref{sec: prob_formu}, the optimal rate region $\mathcal{C}^*$ is characterized by
\begin{align}
\mathcal{C}^* = 
\left\{
\begin{aligned}
&(R_1, R_2,\\
&\quad R_S, R_{S_\Sigma})
\end{aligned}
\middle|
\begin{aligned}
&R_1 \ge \frac{d}{d-s},\\
&R_2 \ge \max\!\left\{\frac{1}{d-s},\frac{1}{K-1}\right\},\\
&R_S \ge \max\!\left\{\frac{1}{d-s},\frac{1}{K-1}\right\}\cdot\frac{\log p}{\log q},\\
&R_{S_\Sigma} \ge \max\!\left\{\frac{d}{d-s},\frac{K-d}{d-s}\right\}\cdot\frac{\log p}{\log q},
\end{aligned}
\right\}\label{eq: rate_region}
\end{align}
where $p$ is a prime\footnote{ Notably, modulo equations over finite fields admit unique solutions and support flexible algebraic operations, owing to the prime field size. In contrast, modulo equations over other groups (or e.g., real-field torus \cite{jaramillovelez2025perfectlyprivateanalogsecureaggregation}) generally do not admit a unique solution, which limits its applicability.} satisfying $p > K(q-1)$.
\end{theorem}
\begin{remark}
When $d=K$ and $s=0$, the protocol is communication-efficient but does not provide resilience to unreliable communication. In this case, the optimal key rate is $(1, \frac{1}{K-1}, \frac{1}{K-1}\frac{\log p}{\log q}, \frac{1}{K-1}\frac{\log p}{\log q})$, due to the relay security constraint. Compared to \cite{zhang2025fundamentallimitshierarchicalsecure}, Theorem~\ref{Theo: 1} additionally incorporates a decoding perspective, which establishes the achievability of the resulting optimal rate region.
\end{remark}

\section{Achievability}
\subsection{Example}
Consider the setup, in which $K=5$, $L=2$, $s=1$, $d=3$. Assume each client $k$ holds $\boldsymbol{\Theta}_k\in \{0, 1, 2\}^2$. Let $p=13$, $S_{\sum}=(Z_1, Z_2, Z_3)$, where $Z_k$ is i.i.d. uniformly distributed over $\mathbb{F}_{13}$. The secret keys are generated by 
\begin{align}
    \begin{bmatrix}
        \mathbf{S}_1\\
        \mathbf{S}_2\\
        \mathbf{S}_3\\
        \mathbf{S}_4\\
        \mathbf{S}_5\\
    \end{bmatrix}
    = \mathbf{G}_S 
    \begin{bmatrix}
        Z_1\\
        Z_2\\
        Z_3\\
    \end{bmatrix}
    =\begin{bmatrix}
        Z_1\\
        Z_2\\
        Z_3\\
        Z_1\oplus 2Z_2\oplus4Z_3\\
        11Z_1\oplus10Z_2\oplus8Z_3\\
    \end{bmatrix},
\end{align}
such that any three secret keys are linearly independent and their sum is zero.
Each client $k$ sends the following messages to relays in $\mathcal{R}_k$:
\begin{align}
&\begin{cases}
&\hspace{-2mm}\mathbf{X}_{4,1}=11\boldsymbol{\Theta}_1(1)\oplus 11 \mathbf{S}_1, \\
&\hspace{-2mm}\mathbf{X}_{5,1}=3\boldsymbol{\Theta}_1(1)\oplus 3\boldsymbol{\Theta}_1(2)\oplus 3\mathbf{S}_1,\\
&\hspace{-2mm}\mathbf{X}_{1,1}=\boldsymbol{\Theta}_1(1) \oplus 10\boldsymbol{\Theta}_1(2)\oplus \mathbf{S}_1. 
\end{cases}\\
&\begin{cases}
&\hspace{-2mm}\mathbf{X}_{5,2}=\boldsymbol{\Theta}_2(1)\oplus 3\boldsymbol{\Theta}_2(2)\oplus \mathbf{S}_2, \\
&\hspace{-2mm}\mathbf{X}_{1,2}=3\boldsymbol{\Theta}_2(1)\oplus 3\boldsymbol{\Theta}_2(2)\oplus 3\mathbf{S}_2,\\
&\hspace{-2mm}\mathbf{X}_{2,2}=2\boldsymbol{\Theta}_2(1) \oplus 2\mathbf{S}_2.
\end{cases}\\
&\begin{cases}
&\hspace{-2mm}\mathbf{X}_{1,3}=6\boldsymbol{\Theta}_3(1)\oplus 6\boldsymbol{\Theta}_3(2)\oplus 6\mathbf{S}_3, \\
&\hspace{-2mm}\mathbf{X}_{2,3}=6\boldsymbol{\Theta}_3(1)\oplus 12\boldsymbol{\Theta}_3(2)\oplus 6\mathbf{S}_3, \\
&\hspace{-2mm}\mathbf{X}_{3,3}=\boldsymbol{\Theta}_3(1)\oplus 3\boldsymbol{\Theta}_3(2) \oplus \mathbf{S}_3.
\end{cases}\\
&\begin{cases}
&\hspace{-2mm}\mathbf{X}_{2,4}= 10\boldsymbol{\Theta}_4(1)\oplus 3\boldsymbol{\Theta}_4(2)
\oplus 10\mathbf{S}_4, \\
&\hspace{-2mm}\mathbf{X}_{3,4}= 11\boldsymbol{\Theta}_4(1)\oplus11 \mathbf{S}_4,\\
&\hspace{-2mm}\mathbf{X}_{4,4}=3\boldsymbol{\Theta}_4(1)\oplus 3\boldsymbol{\Theta}_4(2)\oplus 3\mathbf{S}_4.
\end{cases}\\
&\begin{cases}
&\hspace{-2mm}\mathbf{X}_{3,5}=\boldsymbol{\Theta}_5(1)\oplus 10\boldsymbol{\Theta}_5(2)\oplus \mathbf{S}_5, \\
&\hspace{-2mm}\mathbf{X}_{4,5}=7\boldsymbol{\Theta}_5(1)\oplus 12\boldsymbol{\Theta}_5(2)\oplus 7\mathbf{S}_5,\\
&\hspace{-2mm}\mathbf{X}_{5,5}=6\boldsymbol{\Theta}_5(1)\oplus7 \boldsymbol{\Theta}_5(2)\oplus 6\mathbf{S}_5.
\end{cases}
\end{align}

Subsequently, each relay constructs the following message from the received messages and forwards it to the server.
\begin{align}
\vspace{-2mm}
\mathbf{Y}_1&=\hspace{-1mm}\underset{k\in\mathcal{U}_1}{\oplus}\mathbf{X}_{1,k}\hspace{-1mm}=\boldsymbol{\Theta}_1(1) \oplus 10\boldsymbol{\Theta}_1(2)
\oplus 3\boldsymbol{\Theta}_2(1)\oplus \notag\\
&\hspace{-5mm}3\boldsymbol{\Theta}_2(2)\oplus 6\boldsymbol{\Theta}_3(1)\oplus 6\boldsymbol{\Theta}_3(2)
\oplus \underbrace{\mathbf{S}_1\oplus 3\mathbf{S}_2\oplus 6\mathbf{S}_3}_{S^Y_1}, \\
\mathbf{Y}_2&=\hspace{-1mm}\underset{k\in\mathcal{U}_2}{\oplus} \mathbf{X}_{2,k}
=2\boldsymbol{\Theta}_2(1)
\oplus 6\boldsymbol{\Theta}_3(1)\oplus 12\boldsymbol{\Theta}_3(2)\oplus \notag\\
& 10\boldsymbol{\Theta}_4(1)\oplus 3\boldsymbol{\Theta}_4(2)
\oplus 2\mathbf{S}_2 \oplus 6\mathbf{S}_3 \oplus 10\mathbf{S}_4,\\
\mathbf{Y}_3&=\hspace{-1mm}\underset{k\in\mathcal{U}_3}{\oplus} \mathbf{X}_{3,k}=\boldsymbol{\Theta}_3(1)\oplus 3\boldsymbol{\Theta}_3(2)
\oplus 11\boldsymbol{\Theta}_4(1)
\notag\\
&\oplus \boldsymbol{\Theta}_5(1)\oplus 10\boldsymbol{\Theta}_5(2) \oplus \mathbf{S}_3 \oplus 11\mathbf{S}_4 \oplus\mathbf{S}_5,\\
 \mathbf{Y}_4&=\hspace{-1mm}\underset{k\in\mathcal{U}_4}{\oplus}\mathbf{X}_{4,k}=
 11\boldsymbol{\Theta}_1(1)\oplus 
 3\boldsymbol{\Theta}_4(1)\oplus 3\boldsymbol{\Theta}_4(2) \notag\\
 &\oplus 7\boldsymbol{\Theta}_5(1)\oplus 12\boldsymbol{\Theta}_5(2) \oplus 11\mathbf{S}_1\oplus 3\mathbf{S}_4\oplus 7\mathbf{S}_5,\\
 \mathbf{Y}_5&=\hspace{-1mm}\underset{k\in\mathcal{U}_5}{\oplus} \mathbf{X}_{5,k}=
 3\boldsymbol{\Theta}_1(1)\oplus 3\boldsymbol{\Theta}_1(2)
  \oplus\boldsymbol{\Theta}_2(1)
  \oplus \notag\\
 &\hspace{-5mm}3\boldsymbol{\Theta}_2(2)\oplus
 6\boldsymbol{\Theta}_5(1) \oplus 7\boldsymbol{\Theta}_5(2) \oplus 3\mathbf{S}_1 \oplus \mathbf{S}_2\oplus 6\mathbf{S}_5.
\end{align}
In such a scheme, the combination matrices are $\mathbf{C}_1=\left[\begin{smallmatrix}
0 & 7 & 11 & 6 & 0 \\
0 & 2 & 1 & 7 &9
\end{smallmatrix}\right]$, 
$\mathbf{C}_2=\left[\begin{smallmatrix}
10 & 0 & 6 & 0 & 10 \\
1 &  0 & 7 & 9 & 10
\end{smallmatrix}\right]$, 
$\mathbf{C}_3=\left[\begin{smallmatrix}
9 & 2 & 0 & 11 & 9 \\
2 & 11 & 0 & 11 &11
\end{smallmatrix}\right]$, 
$\mathbf{C}_4=\left[\begin{smallmatrix}
10 & 0 & 6 & 0 & 10 \\
3 & 9 & 6 & 0 &12
\end{smallmatrix}\right]$, 
$\mathbf{C}_5=\left[\begin{smallmatrix}
0 & 7 & 11 & 6 & 0 \\
4 & 7 & 12 & 2 & 0
\end{smallmatrix}\right]$, 
If the server does not see $ \mathbf{Y}_1$, it can compute the sum  according to $\mathbf{C}_1$, i.e.,  
\begin{align}
    7 \mathbf{Y}_2 \oplus 11 \mathbf{Y}_3
    \oplus 6 \mathbf{Y}_4&= \underset{k \in [K]}{\oplus} \boldsymbol{\Theta}_k(1).\\
    2 \mathbf{Y}_2 \oplus  \mathbf{Y}_3
    \oplus 7 \mathbf{Y}_4 \oplus 9 \mathbf{Y}_5
    &=\underset{k \in [K]}{\oplus} \boldsymbol{\Theta}_k(2).
\end{align}
Other cases with one missing $\mathbf{Y}_m$ can be verified similarly.

Each relay $m$ may receives $ \{\mathbf{X}_{m,k}\}_{k\in\mathcal{U}_m} $, in which the $d$ secret keys $ \{\mathbf{S}_k\}_{k\in\mathcal{U}_m} $ are independent over $\mathbb{F}_{13}$, therefore $I\big( \{\boldsymbol{\Theta}_k\}_{k\in[K]} ;\{\mathbf{X}_{m,k}\}_{k\in\mathcal{U}_m} \big)=0$. 
At the server, 
\begin{align}
    \begin{bmatrix}
        S^Y_1\\
        S^Y_2\\
        S^Y_3\\
        S^Y_4\\
        S^Y_5\\
    \end{bmatrix}
    = 
    \begin{bmatrix}
        1&3&6\\
        10&9&7\\
        9&6&1\\
        0&11&3\\
        4&9&9\\
    \end{bmatrix} 
    \begin{bmatrix}
        Z_1\\
        Z_2\\
        Z_3\\
    \end{bmatrix}, 
\end{align}
It can be verified that $\{S^Y_k\}_{k\in[K]}$ can only be eliminated through $\mathcal{C}=\{ \mathbf{C}_1, \cdots, \mathbf{C}_5 \}$, i.e., when $\oplus_{k \in [K]} \boldsymbol{\Theta}_k$ is attained. Hence, server security in \eqref{eq:security_server} is satisfied. 
Moreover, in this scheme, $R_1=\frac{3}{2}$, $R_2=\frac{1}{2}$, $R_S=\frac{1}{2}\log{13}$, $R_{S_{\sum}}=\frac{3}{2}\log 13$, which achieves the optimal rates in \eqref{eq: rate_region}.

\subsection{General Scheme}
\subsubsection{Rationale} The key idea of the general achievable scheme is to exploit the structured aggregation enforced by communication-efficient gradient coding (ComEffGC) \cite{ye2018communication,zhang2025fundamentallimitshierarchicalsecure}. Although the encoding and decoding operations mix the input space across dimensions, the linear mapping from the inputs $x_k\in \mathbb{F}_p^L$ to the desired sum $\oplus_{k\in[K]} x_k\in \mathbb{F}_p^L$ remains fixed and independent of the transmitted values $x_k$.  
This invariance enables us to build security on top of a communication-efficient and straggler-resilient aggregation scheme. 
In particular, when $\tilde{L}\leq d-s$, a $(K,d-s,s)$-ComEffGC scheme can be specified by a collection of linear encoding functions $\mathcal{G}$,
\begin{align}
    \mathcal{G}&=\left\{ g_m\left( \{\boldsymbol{\Theta}_k(l)\}_{l\in [\Tilde{L}], k\in \mathcal{U}_m }   \right) \right\}_{m\in [K]},
    \label{eq: gm}
\end{align}
where each encoding function $g_m(\cdot): \mathbb{F}_p^{d\Tilde{L}}\rightarrow \mathbb{F}_p $, and a set of combination matrices $\mathcal{C}=\{ \mathbf{C}_1, \cdots, \mathbf{C}_F \}$,  
where each $\mathbf{C}_f=\left[c_{\Tilde{l}, m, f}\right]_{\Tilde{L}\times K}\in \mathbb{F}_p^{\Tilde{L}\times K}$ containing $s$ zero columns corresponds to a possible $s$-straggler pattern such that  
\begin{align}
   &\Bigg\{\hspace{-0.75mm}\underset{k \in [K]}{\oplus}\hspace{-1.5mm} c_{\Tilde{l}, m, f} g_m\hspace{-0.25mm}\left( \{\boldsymbol{\Theta}_k(l)\}_{l\in [\Tilde{L}], k\in \mathcal{U}_m }   \right)  \hspace{-1.5mm}\Bigg\}_{\Tilde{l}\in [\Tilde{L}]} 
   \hspace{-2mm} =\underset{k \in [K]}{\oplus}\boldsymbol{\Theta}_k .
   \label{eq:server_comp}
\end{align}
Since $g_m(\cdot)$ is a linear function, \eqref{eq: gm} can be decomposed into
\begin{align}
&g_m\left( \{\boldsymbol{\Theta}_k(l)\}_{l\in [L], k\in \mathcal{U}_m }   \right) \notag\\
&\hspace{1.5cm}=h_m\left( \left\{ f_{m,k}\left(\{\boldsymbol{\Theta}_k(l)\}_{l\in [L]}\right)\right\}_{k\in \mathcal{U}_m} \right),
\end{align}
where both $h_m(\cdot)$ and $f_{m,k}(\cdot)$ are linear functions over $\mathbb{F}_p$. This compositional structure forms the basis of our general SA scheme. 

\subsubsection{General Scheme} \label{sec: general scheme}
When $d-s< K$, each client $k$ divides its message $\boldsymbol{\Theta}_k$ into $L_X=\frac{L}{d-s}$ segments, i.e., 
\begin{align}
\boldsymbol{\Theta}_k^{(l_X)}=\left\{ \boldsymbol{\Theta}_k(l) \right\}_{l=(l_X-1)(d-s)+1}^{l_X(d-s)}, l_X=1, \cdots, L_X. 
\end{align}
Each client holds a secret key $\mathbf{S}_k=[S_{k,1}, \cdots, S_{k, L_X}]$, and $S_{k, l_X}$ is independent of each other for $\forall l_X$. To ensure correctness of the global model, $\{\mathbf{S}_k\}_{k\in[K]}$ should satisfy $\oplus_{k \in [K]} \mathbf{S}_k=\mathbf{0}$. Assume the secret keys are generated by
\vspace{-2mm}
\begin{align}
\begin{bmatrix}
     S_{1,l_X}\\
     \vdots\\
     S_{K,l_X}
\end{bmatrix}
= \mathbf{G}_S
\begin{bmatrix}
     Z_{(l_X-1)\frac{L_{\sum}}{L_X}+1}\\
     \vdots\\
     Z_{(l_X-1)\frac{L_{\sum}}{L_X}+\frac{L_{\sum}}{L_X}}
\end{bmatrix}
,\forall l_X.
\label{eq: key_construction}
\end{align}
The randomness vector $\mathbf{S}_{\sum}=(Z_1, Z_2, \cdots, Z_{L_{\sum}})$ consists of $L_{\sum}=\max\{d, K-d\}$ i.i.d. symbols, each drawn uniformly from $\mathbb{F}_p$. To guarantee relay security, we require that any $d$ rows in $\mathbf{G}_S$ have rank $d$. To guarantee server security, we require that any $K-d$ rows in $\mathbf{G}_S$ have rank $K-d$. So $ \mathrm{Rank}(\mathbf{G}_S)=\max\{ d, K-d \}$.  A construction of such $\mathbf{G}_S$ can be found in \cite{zhang2025fundamentallimitshierarchicalsecure}. 

Let $\mathbf{S}_k^{(l_X)}=[S_{k, l_X}, \mathbf{0}_{1\times L_X}]^\top$\footnote{$S_{k,l_X}$ can be placed in other entries, depending on the specific GC code.}, each client $k$ sends $\mathbf{X}_{m, k}$ to relay $m\in\mathcal{R}_k$, where
\begin{align}
    &\mathbf{X}_{m, k}(l_X)
    \hspace{-0.5mm}
    =\hspace{-0.5mm} f_{m,k}\hspace{-1mm}\left(\hspace{-1mm}\left\{ \boldsymbol{\Theta}_k^{(l_X)}(l)\oplus \mathbf{S}_k^{(l_X)}(l)\right\}_{l=(l_X-1)(d-s)+1}^{l_X(d-s)} \hspace{-0.5mm}\right), \notag\\
    &\hspace{6cm}\forall l_X\in[L_X]. 
    \label{eq: client-to-relay message}
\end{align}
In this scheme, $\mathbf{S}_{m,k}=\mathbf{S}_k$. The secret key is reused for transmissions to different relays without sacrificing security. 

If relay $m$ receives all messages from $\mathcal{U}_m$,  it computes 
\begin{align}
&\mathbf{Y}_m(l_X) = h_m( \{\mathbf{X}_{m, k}(l_X) \}_{k\in\mathcal{U}_m}), \forall m, \forall l_X.
    \label{eq: encode_Y_scheme}
\end{align}
Then, it transmits $\mathbf{Y}_m$ to the server. If the server receives $\{\mathbf{Y}_m(l_X)\}_{m\in\mathcal{V}_2} $ and that $\lvert \mathcal{V}_2 \rvert \geq K-s$, the server can combine $\{\mathbf{Y}_m(l_X)\}_{m\in\mathcal{V}_2} $ as in \eqref{eq:server_comp},  and acquire 
\begin{align}
     \underset{k \in [K]}{\oplus} \left(\boldsymbol{\Theta}_k^{(l_X)}\oplus \mathbf{S}_k^{(l_X)} \right) \hspace{-1mm}=\hspace{-2mm}\underset{k \in [K]}{\oplus} \boldsymbol{\Theta}_k^{(l_X)}, \forall l_X.
    \label{eq: server_agg}
\end{align}
Hence, the server can recover the exact sum.

\subsubsection{Achievability Proof}
The correctness of the server aggregation was proved in the previous section. We now proceed to analyze other performance metrics. 

\textit{Relay Security:} Since $ \mathrm{Rank}(\mathbf{G}_S)=\max\{ d, K-d \}$, each relay $m$ can receive $d$ independent secret keys, we have
\begin{align}
    &I\big( \{\boldsymbol{\Theta}_k\}_{k\in[K]} ;\{\mathbf{X}_{m,k}\}_{k\in\mathcal{U}_m} \big)\notag\\
    &= H\left(\{\mathbf{X}_{m,k}\}_{k\in\mathcal{U}_m})-H(\{\mathbf{X}_{m,k}\}_{k\in\mathcal{U}_m}\vert \{\boldsymbol{\Theta}_k\}_{k\in[K]} \right)   \notag\\
    &\overset{\eqref{eq: key_construction}}{=} H\left(\{\mathbf{X}_{m,k}\}_{k\in\mathcal{U}_m})-H(\{\mathbf{S}_{k}\}_{k\in\mathcal{U}_m} \right)   \notag\\
    &\leq d L_X\log p-d L_X\log p \notag \\
    &=0.
    \label{eq: Relay Security}
\end{align}

\textit{Server Security:} At the server, the global model can be formed with any $K-s$ relay computations $\mathbf{Y}_m$, due to linearity, it is unlikely to exist other combination beyond $\mathcal{C}$ that can result in the sum of secret keys and completely cancel secret keys as in \eqref{eq: server_agg} regardless of the cardinality of $\mathcal{V}_2$. We only need to examine whether $\mathbf{G}_S$ permits additional matrix manipulations that would cancel the secret keys. 
Let $\mathbf{S}^Y_k(l_Y)$ denote the linear functions of secret keys in $\mathbf{Y}_m(l_Y)$. To achieve full cancellation of the secret keys when computing the exact sum of length $L_Y$ with any $K-s$ relay computations $\mathbf{Y}_m(l_Y)$, there must exist $K-(d-s)-s+1=K-d+1$ functions $\mathbf{S}^Y_k(l_Y)$ that cancel out.
If $d>K-d$, i.e., $d>\frac{K}{2}$,  $\mathrm{Rank}(\mathbf{G}_S)=d$, at least $d+1> K-d+1$ independent containing $\mathbf{S}^Y_k(l_Y)$ are required to eliminate randomness, which contradicts the previous discussion. If $d\leq K-d$, i.e., $d\leq\frac{K}{2}$, then $\mathrm{Rank}(\mathbf{G}_S)=K-d$. For $\forall\mathcal{V}_2: \lvert\mathcal{V}_2\rvert=K-s$, which implies the null space coincides with the column span of the corresponding $\mathbf{C}_f$ in $\mathcal{C}$. 
Formally, for $\forall\mathcal{V}_2: \lvert\mathcal{V}_2\rvert<K-s$, 
\begin{align}
&I\bigg( \{\boldsymbol{\Theta}_k\}_{k\in[K]} ; \left\{\mathbf{Y}_m\right\}_{k\in\mathcal{V}_2} \bigg\vert \underset{k \in [K]}{\oplus} \boldsymbol{\Theta}_k \bigg) \notag\\
&=H\left( \hspace{-1mm}\left\{\mathbf{Y}_m\right\}_{k\in\mathcal{V}_2}\hspace{-0.5mm}\bigg\vert \underset{k \in [K]}{\oplus} \boldsymbol{\Theta}_k\hspace{-1mm}\right)
\hspace{-0.5mm}-\hspace{-0.5mm}H\left( \left\{\mathbf{Y}_m\right\}_{k\in\mathcal{V}_2}\hspace{-1mm}\vert \{\boldsymbol{\Theta}_k\}_{k\in[K]} \right) \notag\\
&\leq H\left( \left\{\mathbf{Y}_m\right\}_{k\in\mathcal{V}_2}\vert \{\boldsymbol{\Theta}_k\}_{k\in[K]}\right)
-H\left( \left\{\mathbf{S}^Y_k\right\}_{k\in\mathcal{V}_2} \right) \notag\\
&=H\left( \left\{\mathbf{S}^Y_k\right\}_{k\in\mathcal{V}_2}\right)
-H\left( \left\{\mathbf{S}^Y_k\right\}_{k\in\mathcal{V}_2} \right) \notag\\
&=0 .  \label{eq: server Security}
\end{align}
For $\forall\mathcal{V}_2: \lvert\mathcal{V}_2\rvert\geq K-s$, the proof is similar to \cite[(66)]{zhang2025fundamentallimitshierarchicalsecure}.

\textit{Metrics:} The rates of the proposed general scheme are calculated as follows. 
\begin{itemize}
    \item $R_1:$ Each client $k$ send $\mathbf{X}_{m, k}$ of length $L_X$ to $d$ relays in $\mathcal{R}_k$. As a result of \eqref{eq: key_construction} and \eqref{eq: client-to-relay message}, each $\mathbf{X}_{m, k}$ are i.i.d. distributed over $\mathbb{F}_p^{L_X}$. So $H(\mathbf{X}_{m, k})=L_X \log p$, $R_1=\frac{d H(\mathbf{X}_{m, k})}{L\log p}=\frac{d}{d-s}$.   
    \item $R_2:$  Following \eqref{eq: encode_Y_scheme}, $L_Y=L_X$ and that $\mathbf{Y}_m(l_X)$, $l_X\in [L_X]$, are i.i.d. uniformly distributed in $\mathbb{F}_q^{L_Y}$. We have $H(\mathbf{Y}_m)=L_Y\log p$, $R_2=\frac{KH(\mathbf{Y}_m)}{KL\log p}=\frac{1}{d-s}$.  
    \item $R_S:$ Following \eqref{eq: key_construction}, each $\mathbf{S}_k$ contains $L_X$ symbols i.i.d. uniformly distributed over $\mathbb{F}_p^{L_X}$, so $H(\mathbf{S}_k)=L_X \log p$, $R_S=\frac{1}{d-s}\frac{\log p}{\log q}$. 
    \item $R_{S_{\sum}}:$ Following \eqref{eq: key_construction} and the rank requirements of $\mathbf{G}_S$ posed by relay and server security, $\frac{L_{\sum}}{L_X}=\max\{d, K-d\}$. So $H(\mathbf{S}_{\sum})=L_X \cdot \max\{d, K-d\} \log p$, and $R_{S_{\sum}}=\max\!\left\{\frac{d}{d-s},\frac{K-d}{d-s}\right\}\cdot\frac{\log p}{\log q}$.
\end{itemize}

When $d-s= K$, i.e., $d=K$ and $s=0$, the $(K,K-1,0)$-ComEffGC codes can be adopted, and the transmission scheme remains the same. The security check in \eqref{eq: Relay Security} and \eqref{eq: server Security} remains valid. $R_1=\frac{1}{K-1}$, $R_2=\frac{1}{K-1}$, $R_S=\frac{1}{K-1}\frac{\log p}{\log q}$, $R_{S_{\sum}}=\max\left\{\frac{d}{K-1},\frac{K-d}{K-1}\right\}\cdot\frac{\log p}{\log q}$.

\section{Converse}
\subsubsection{Proof of $R_1\geq \frac{d}{d-s}$
}\label{sec: proof of R1}
Regardless of security constraints or the specific encoding and decoding scheme, correctly computing the entry-wise sum of $K$ inputs, each with $L$ DoF, requires access to all $L$ DoF from every input. Each client $k$ must make $L$ DoF available collectively to its associated relays, even with $s$ communication failures. So any $d-s$ messages $\mathbf{X}_{m,k}$, $m\in\mathcal{R}_k$, and $\mathbf{X}_{k,k}$ should together carry a total of $L$ DoF. Due to the network symmetry, $L_X\geq\frac{L}{d-s}$.

The relay security in \eqref{eq:security_relay} further implies that for each individual message $\mathbf{X}_{m,k}$, 
$I(\boldsymbol{\Theta}_k; \mathbf{X}_{m,k})=0$, $\forall k\in\mathcal{U}_m$.  Combined with the fact that $\mathbf{X}_{m,k}\in \mathbb{F}_p^{L_X}$ is a linear function in \eqref{eq:encode}, $\mathbf{S}_{m,k}$ must be i.i.d. uniformly distributed in $\mathbb{F}_p^{L_S}$ and that $H(\mathbf{S}_{m,k})\geq H(\mathbf{X}_{m,k})=L_X \log p$. 
Hence, $R_1=\frac{\sum_{m\in\mathcal{R}_k}H(\mathbf{X}_{m,k})}{L\log p}
\geq \frac{d H(\mathbf{X}_{m,k})}{L\log p}= \frac{d L_X \log p}{L\log p} \geq \frac{d}{d-s}$.

\subsubsection{Proof of $R_2\geq \max\left\{\frac{1}{d-s},\frac{1}{K-1} \right\}$}  \label{sec: proof_R2}
First, due to the linear relation in \eqref{eq: encode_Y}, it holds that
\begin{align}
    H(\mathbf{Y}_m)
    &\geq H \left(\mathbf{Y}_m\mid \{ \mathbf{X}_{m,k} \}_{k\in \mathcal{U}_m\setminus \{k\} }\right) \notag\\
    &\overset{\eqref{eq: encode_Y}}{=}H(\mathbf{X}_{m,k}).
\end{align}
Hence, 
$R_2\triangleq\frac{\sum_{m\in[K]} H(\mathbf{Y}_m)}{KL\log p}\geq \frac{1}{d-s}$. Consider network symmetry, we additionally get $L_Y\geq\frac{L}{d-s}$. 

Next, we prove that $R_2=\frac{1}{K}$ is not attainable. When $d-s=K$, i.e., $d=K$, $s=0$, the server requires all $\left\{\mathbf{Y}_m\right\}_{m\in[K]}$ to be able to compute the sum correctly. If $R_2=\frac{1}{d-s+1}=\frac{1}{K}$, the server sees all $\left\{\mathbf{Y}_m\right\}_{m\in[K]}$ with a total of $\frac{L}{K}\cdot K=L$ DoF, and wish to compute the exact sum.  
Since each $\mathbf{Y}_m$ is a linear function of
$\{\boldsymbol{\Theta}_i(l)\}_{i \in [K],\, l \in [L]}$ and that the total number of DoF observed by the server is $L$, the exact recovery of
the sum with zero error
is possible only if the collection $\{\mathbf{Y}_m\}_{k\in[K]}$ consists of $L$
linearly independent linear combinations of the desired sums
$\left\{ \oplus_{k \in [K]} \boldsymbol{\Theta}_k(l)\right\}_{l=1}^L$, i.e., $\mathbf{Y}_m=\boldsymbol{a}_m\left(\oplus_{k \in [K]} \boldsymbol{\Theta}_k\right)$. Note that any $\mathbf{Y}_m$ must not contain privacy randomness, since the cancellation of the randomness will consume at least $1$ degree of freedom and make the exact recovery impossible.    
In this case, the server security is met:
\vspace{-2mm}
\begin{align}
&I\bigg( \{\boldsymbol{\Theta}_k\}_{k\in[K]} ; \left\{\mathbf{Y}_m\right\}_{m\in [K]} \bigg\vert \underset{k \in [K]}{\oplus} \boldsymbol{\Theta}_k \bigg) \notag\\
=&I\bigg( \{\boldsymbol{\Theta}_k\}_{k\in[K]} ; \left\{\underset{k \in [K]}{\oplus} \boldsymbol{\Theta}_k(l)\right\}_{l=1}^L \bigg\vert \underset{k \in [K]}{\oplus} \boldsymbol{\Theta}_k \bigg) \notag\\
=&0.
\end{align}

\vspace{-3mm}
\noindent
However, the relay security is not met, as $R_2=\frac{1}{K}$ requires that each relay can completely cancel secret keys it receives, violating relay security constraint.  Since that $\mathbf{Y}_m$ is generated from $\{\mathbf{X}_{m, k} \}_{k\in\mathcal{U}_m}$, it holds that
\vspace{-2mm}
\begin{align}
&I\big( \{\boldsymbol{\Theta}_k\}_{k\in[K]} ;\{\mathbf{X}_{m,k}\}_{k\in\mathcal{U}_m} \big) \notag\\
=&I\bigg( \{\boldsymbol{\Theta}_k\}_{k\in[K]} ;\{\mathbf{X}_{m,k}\}_{k\in\mathcal{U}_m}, \boldsymbol{a}_m\left(\underset{k \neq m}{\oplus} \boldsymbol{\Theta}_k\right) \bigg) \notag\\
\geq&I\bigg( \{\boldsymbol{\Theta}_k\}_{k\in[K]} ; \boldsymbol{a}_m\left(\underset{k \neq m}{\oplus} \boldsymbol{\Theta}_k\right) \bigg) >0.
\end{align}


\vspace{-3mm}
\subsubsection{Proof of $R_S\geq  \max\left\{\frac{1}{d-s},\frac{1}{K-1} \right\}\cdot \frac{\log p}{\log q}$}
Following the proof in Section \ref{sec: proof of R1}, we directly arrive at $R_S=\frac{H(\mathbf{S}_k)}{L\log q}\geq\frac{H(\mathbf{S}_{m,k})}{L\log q}\geq\frac{1}{d-s} \frac{\log p}{\log q}$. 

Next, we prove $R_S\neq\frac{1}{K}\frac{\log p}{\log q}$. Due to the correctness of recovery in \eqref{eq:correctness1}, the secret keys $\{\mathbf{S}_{k}\}_{k\in[K]}$ must be able to cancel out, i.e., there exists a function $c(\cdot)$ such that 
\begin{align}
    c(\{\mathbf{S}_{k}\}_{ k\in[K]})=\mathbf{0}.
    \label{eq: containt_security}
\end{align}
When $d=K$, $s=0$, $R_S=\frac{1}{K}\frac{\log p}{\log q}$, each relay hears from all clients, i.e., each relay can receive $\mathbf{X}_{m,k}$, $k\in[K]$, masked by $\{\mathbf{S}_{k}\}_{k\in[K]}$. 
Due to the linear process in the network and \eqref{eq: containt_security}, which is posed by the correctness condition, the relay $m$ must be able to cancel the secret keys completely through the linear function $c_m'(\cdot)$. Then, it follows that
\begin{align}
&I\big( \{\boldsymbol{\Theta}_k\}_{k\in[K]} ;\{\mathbf{X}_{m,k}\}_{k\in[K]}\big) \notag\\
=&I\big( \{\boldsymbol{\Theta}_k\}_{k\in[K]} ;c_m'(f(\{\boldsymbol{\Theta}_k(l)\}_{k\in[K], l\in[L]}, \{\mathbf{S}_{k}\}_{k\in[K]})) \big) \notag\\
=&I\big( \{\boldsymbol{\Theta}_k\}_{k\in[K]} ;c_m''( \{\boldsymbol{\Theta}_k(l)\}_{k\in[K], l\in[L]}) \big)>0,
\end{align}
\noindent which violates the peer security constraint in \eqref{eq:security_relay}.

\subsubsection{Lower bound for $R_{S_{\sum}}\geq \max\left\{\frac{d}{d-s}, \frac{K-d}{d-s}  \right\}\cdot \frac{\log p}{\log q}$}
Each relay observes $d$ messages $\{\mathbf{X}_{m,k}\}_{k\in\mathcal{U}_m}$ from $d$ clients in $\mathcal{U}_m$, each $\mathbf{X}_{m,k}$ of length $L_X$. To maintain relay security in \eqref{eq:security_relay}, all $d L_X$ entries in $\{\mathbf{X}_{m,k}\}_{k\in\mathcal{U}_m}$ must be independent. This further requires at least $d L_X$ independent randomness in $\mathbf{S}_{\sum}$ such that all $d L_X$ entries in $\{\mathbf{S}_{m,k}\}_{k\in\mathcal{U}_m}$ are independent. Due to linear generation of the secret keys using $\mathbf{S}_{\sum}$ and the i.i.d. uniform distribution of $\{\mathbf{S}_{m,k}\}_{k\in\mathcal{U}_m}$ in $\mathbb{F}_p^{L_S}$ (discussed in Section \ref{sec: proof of R1}), each independent random variable $Z_i$ in $\mathbf{S}_{\sum}$ must also be i.i.d. uniformly distributed in $\mathbb{F}_p^{L_S}$. Hence, $H(\mathbf{S}_{\sum})\geq dL_S\log p\geq dL_X\log p\geq \frac{d}{d-s}L\log p$. By the definition of $R_{S_{\sum}}$ in \eqref{eq: rate_secret}, we directly arrive at $R_{S_{\sum}}\geq \frac{d}{d-s}\frac{\log p}{\log q}$.  

Next, we prove $R_{S_{\sum}}\geq \frac{K-d}{d-s} \frac{\log p}{\log q}$. 
For any $\mathcal{V}_2: \lvert \mathcal{V}_2 \rvert \geq K-s$, $\{\mathbf{Y}_m\}_{k\in \mathcal{V}_2}$ can recover the exact sum with length $L$. To avoid unintended information leakage, any linear combination leading to the key cancellation should correspond to the linear combination that yields the intended sum.
There can only be exactly $L$ linear combinations of $\{\mathbf{Y}_m\}_{k\in \mathcal{V}_2}$ leading to key cancellation. To make this possible, at least $L_{\sum}\geq (K-s) L_Y-L$ independent randomness are required for $\{\mathbf{Y}_m\}_{k\in \mathcal{V}_2}$. If $L_{\sum}< (K-s) L_Y-L$, there exist more key-cancellation combinations than the $L$ linear combinations associated with valid sum computations, leading to information leakage. Hence, $L_{\sum}\geq (\frac{K-s}{d-s}-1) L= \frac{K-d}{d-s}L$. Then, it can be derived that $R_{S_{\sum}}\geq \frac{K-d}{d-s} \frac{\log p}{\log q}$.

\vspace{-0.5em}
\section{Conclusion}
\vspace{-2mm}
This work investigates resilient and efficient HSA under unreliable communication. We characterize the optimal rate region and propose an optimal scheme, supported by matching achievability and converse proofs. Moreover, by guaranteeing a one-to-one correspondence between finite-field and real-field summations, this work takes a crucial step toward bridging information-theoretic secure aggregation with practical learning settings. 

\bibliographystyle{IEEEtran.bst}
\bibliography{IEEEabrv,ref}

\end{document}